\documentstyle[11pt,aspconf]{article}
\begin{document}
\title{"NOT TOO OLD" METAL DEFICIENT STELLAR POPULATIONS:
THE CASE OF METALLICITY Z=0.00001.}
\author{S. Cassisi$^1$,$^2$, M. Castellani$^3$ and V. Castellani$^1$,$^4$}
%
%
\affil     {$^1$ Osservatorio Astronomico Collurania,Via Mentore Maggini,I-64100 Teramo, Italy \\
            $^2$ Universit\'a de l'Aquila, Dipartimento di Fisica, Via Vetoio,I-67010 L'Aquila, Italy \\
            $^3$ Universit\'a "La Sapienza", Istituto Astronomico, Via Lancisi 29,I-00161 Roma, Italy \\
            $^4$ Universit\'a di Pisa, Dipartimento di Fisica, Piazza Torricelli 2,I-56100 Pisa, Italy}
            
%
%
%
%
\begin{abstract}
We investigate the evolution of
metal deficient stellar structures, presenting 
H-burning isochrones covering cluster ages from 800 Myr to 7 Gyr.
Evolutionary evidences for 
selection effects in the metallicity distribution of 
very metal poor H-burning red giants are reported.
The evolution of stars during central and shell He burning is further
investigated, discussing the occurrence of He burning pulsators as a
function of cluster age.

\end{abstract}
%
%
%
\section {Introduction}
\par
Since the pioneering work by Bond (1970), in the past few decades 
much observational
effort has been devoted to searching for very metal poor stars. 
As a consequence, one is dealing with 
an increasing evidence for metal deficient stars membering the 
Galactic Halo (see, e.g., Molaro \& Castelli 1990,
 Primas et al. 1994, Sneden et al. 1994), renewing the interest in
theoretical constraints concerning the evolution of similar very metal
poor objects. 
As matter of fact, even if the approach to the evolution of 
metal deficient stars dates to the early seventies,
this argument is still open to investigation.
In a preliminary paper Cassisi \& Castellani (1993)
presented  a rather extensive 
investigation of the theoretical scenario concerning
 these peculiar stellar objects.
However, their analysis was mainly devoted to the study of the
 \lq{Red Giant Phase Transition}\rq\ (see below)
in low mass stars as well as to the determination of the lower mass limit
 ($M^{up}$) for quiet carbon ignition in more
massive stars. 
This scenario has been recently improved by Cassisi et al. (1995; Paper I) 
who investigated 
H and He burning 
phases for low mass, metal deficient stars, presenting theoretical isochrones
for ages in the range 7 - 15 billion of years and
discussing the evolutionary expectations for  RR-Lyrae stars.
\par
According to these results, one finds that old Population III and Population II
stars have been rather extensively investigated in the literature.
However, the evidence is increasing for dwarfs spheroidals being metal
poor systems, but \lq{not too old}\rq, i.e., not as old as galactic
globular clusters are.
Such an evidence has recently suggested the opportunity to extend to
larger masses evolutionary computations 
concerning metal poor stellar structures. Accordingly, 
Castellani \& Degl'Innocenti (1995) have
discussed the evolutionary behavior of stars up to about $2M_\odot$,
for the two selected choices on the amount of heavy elements: 
$ Z= 10^{-4} - 4\cdot10^{-4}$,
extending in such a way previous investigations of metal poor stars to 
cluster ages lower than 1 Gyr.
However, suggestions have been advanced 
for the occurrence in 
dwarf spheroidal of a not negligible spread of metallicities, with
the possible occurrence of stars with even lower values of Z. On this basis,
Caputo \& Degl'Innocenti (1995) have recently speculated about the
possible occurrence of metal deficient He burning pulsators.
Due to the lack of 
investigation on the evolutionary
scenario concerning similar metal deficient, but not too old,
 stellar systems, it is obviously interesting to extend to lower
metallicities the evolutionary scenario presented in the above referred
literature.
\par
This paper will present the results of such an investigations,
as performed exploring the evolutionary behavior of stars with
$Z= 10^{-5}$, a metallicity value adopted as suitable lower limit
for the amount of heavy elements in dwarf spheroidal systems.
In section
2 we will discuss the H burning evolution for the given 
choice on the stellar metallicity. Starting from these
results, section 3 will deal with the results concerning He burning phases. 
A final discussion will close the paper.
\section{The H burning phase in metal deficient stars.}
\par
In order to extend the theoretical scenario
concerning stars with $Z=10^{-5}$ 
down to cluster ages
of about 1 Gyr, evolutionary tracks already presented in Paper I 
for the quoted metallicity have been implemented
with new tracks for suitable choices of the stellar masses.
All the computations have been performed adopting a 
cosmological abundance of He as given by $Y=0.23$.
\par
As well known, the modality of He ignition significantly
depends on the total mass of the evolving giant. For each given
stellar population (i.e., for each assumed value of $Y$ and $Z$),
one may define a critical mass $M_{HeF}$ as 
the upper mass limits for stars  
 experiencing strong electron degeneracy 
of the He core during the H shell 
 burning phase and - thus - igniting He through one 
or more violent He flashes.
One finds that evolutionary features of red giant
(RG) stars with masses around $M_{HeF}$ change in a remarkable
way in a range of only few tenths of solar mass, 
an occurrence already known as
\lq{\sl Red
Giant Branch Transition}\rq\ (RGT) 
(see Sweigart, Greggio \& Renzini 1989, hereafter SGR,
Sweigart, Greggio \& Renzini 1990,
Castellani et al. 1992).
\par
\begin{figure}
\epsscale{.60}
\plotone{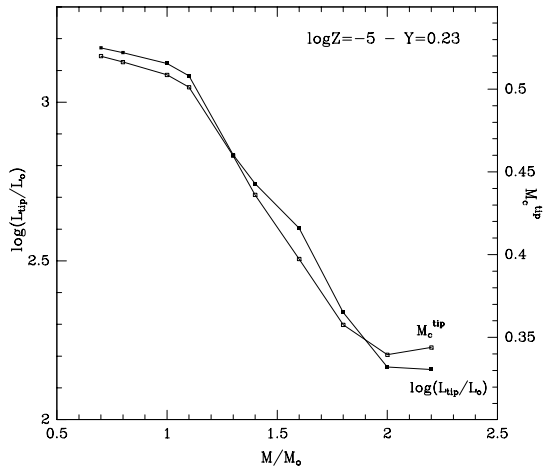}
\caption{The luminosity of the RGB tip and the mass of
the helium core at the helium ignition versus the total star mass
when Z=$10^{-5}$.}
\end{figure}

The behavior of our Z=$10^{-5}$ models through the transition is
shown in
figure 1, which shows the dependence of $M^{tip}_c$ 
(the mass of the He core at the
He ignition) and $L_{tip}$ (the star luminosity at the tip of RGB)  
on the stellar mass.
According to Sweigart \& Gross (1978) and SGR, the onset
of the helium flash has been taken at the model
where the contribution of $3\alpha$ reactions
to the energetic reaches $100L_\odot$; for structures which
quietly ignite helium, the He ignition has been alternatively 
fixed at the first appearance
of a convective core.
The sudden variation of $L_{tip}$ 
around M=$1.5M_{\odot}$ indicates
that this stellar mass is near 
the transition between low mass stars developing full degenerate Helium cores
and more massive structures where electron degeneracy is progressively 
removed.
As in previous investigations, if one defines the critical mass $M_{HeF}$
as the mass of the star having at the He ignition a He core mass equal to
the average value between the He core of fully degenerated structures
and the absolute minimum in $M^{tip}_c$, when Z=$10^{-5}$
one finds  
$M_{HeF}$ of the order of $1.45M_\odot$.
Table 1 reports selected evolutionary parameters for all the 
computed models, allowing a quantitative inspection of the RGB transition.
\par

\begin{figure}
\epsscale{.60}
\plotone{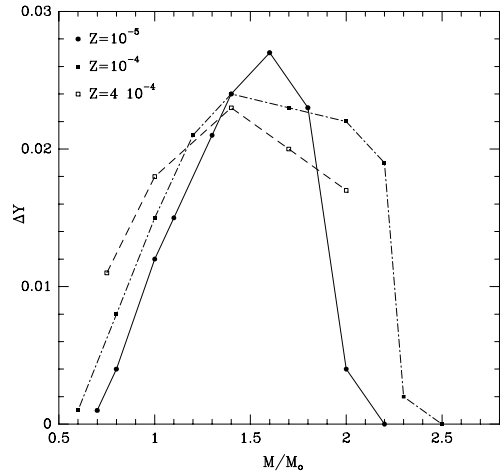}
\caption{The amount of extra-helium brought to the surface by the first dredge up
 for all the computed models. For the sake of comparison, the results concerning
two different assumptions  about the stellar metallicity are reported.}
\end{figure}
Figure 2 compares the amount of extrahelium brought to the surface by the first
dredge up with similar data but for the larger metallicities investigated
in Paper I. 
As already discussed in Castellani \& Degl'Innocenti (1995), 
for each given metallicity one finds
a stellar mass
separating the regime
of low mass stars where $\Delta{Y}$ increases when the stellar mass
is increased from more massive stars with opposite behavior.
Such an occurrence as well as the dependence of $\Delta{Y}$ on the
star metallicity can be easily understood in terms of the 
discussion given by Castellani \& Degl'Innocenti (1995). 
\par
\begin{figure}
\epsscale{.60}
\plotone{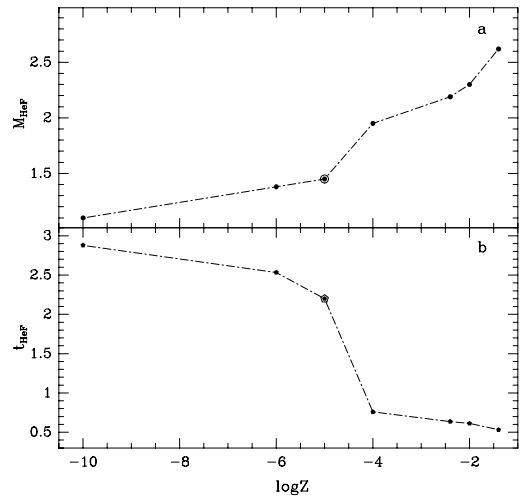}
\caption{The critical mass $M_{HeF}$ (in solar mass) (a) and 
the age (in Gyrs) of a cluster with $M_{HeF}$ at the He ignition (b)
versus the global amount of heavy elements.}
\end{figure}
%
%
%

\begin{deluxetable}{ccccc}
\scriptsize
\tablecaption{Selected evolutionary parameters at the He ignition: 1) the star mass,
2) the stellar luminosity at the tip of the RGB, 3) the mass (in solar unit) of the He core, 4) the amount
of extra-helium brought at the surface by the first dredge up and 5) the age (in Gyrs)
of the star.}
\tablehead{
\colhead{$M/M_{\odot}$} & \colhead{$log(L/L_{\odot})_{tip}$} & 
\colhead{$M_c^{tip}$} & \colhead{$\Delta Y$} & \colhead{$t_{HeF}$}}
\startdata
1.1 &3.083 &0.501 &0.015 &5.04 \nl
1.3 &2.835 &0.460 &0.021 &2.89 \nl
1.4 &2.743 &0.436 &0.024 &2.27 \nl
1.6 &2.603 &0.397 &0.027 &1.47 \nl
1.8 &2.338 &0.358 &0.023 &1.01 \nl
2.0 &2.166 &0.339 &0.004 &0.72 \nl
2.2 &2.158 &0.344 &0.000 &0.53 \nl
\enddata
\end{deluxetable}
%
%
%
Figure 3a shows the dependence of the critical mass $M_{HeF}$ 
on star metallicity. 
In this figure, present results have been implemented 
with similar data given by Cassisi \& Castellani (1993) 
or by SGR for lower or larger metallicities, respectively.
The dependence of $M_{HeF}$ on Z has been already discussed (see Cassisi and Castellani 1993)
and this discussion will not be repeated here.
As a relevant point figure 3b discloses the dependence
on the metallicity of
the cluster age at the Helium ignition in stars with mass M=$M_{HeF}$.
It appears that when Z=$10^{-5}$ the transition requires ages of the order of about 2.2 Gyr,
i.e., a much larger age than for the Z=$10^{-4}$ case. As a consequence,
in a dwarf galaxy with star metallicities ranging from
Z=$10^{-5}$ to Z=$10^{-4}$ and
ages around 1 Gyr, the red giant branch is expected to be populated 
by the more metal rich stars only.
As a result, one finds that
the distribution of
metallicity of RG cannot taken in all cases as a bona fide indicator of the
distribution of star metallicity in \lq{not-too-old}\rq\ metal poor
systems.
\par
Evolutionary models, as computed for the case Z=$10^{-5}$ allow us to
extend toward lower ages the set of isochrones presented in Paper I.
This is shown in figure 4 where we report selected isochrones for H burning
stellar structures covering cluster ages from 0.8 to 7Gyrs.
\begin{figure}
\epsscale{.60}
\plotone{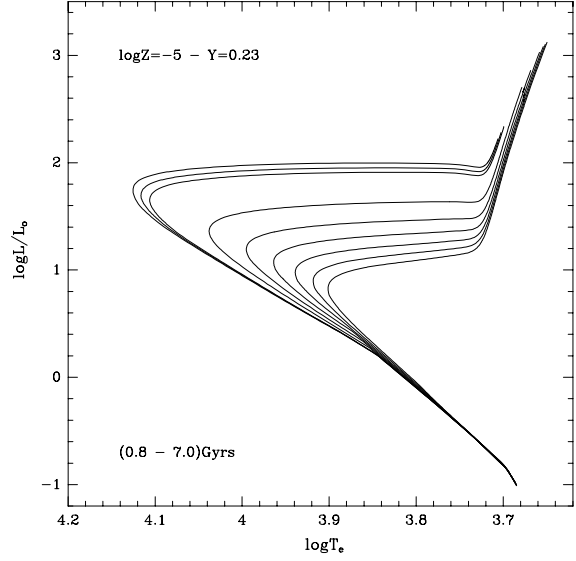}
\caption{Cluster isochrones for H burning phases and for the labeled range
of ages. The interval is 100 Myr for ages lower than 1 Gyr and 1 Gyr for 
larger ages.}
\end{figure}
\section{The evolution along the He burning phase.}

Evolutionary data for H burning stars, as given in the previous section,
allow to investigate the evolution of stellar
structures along the He burning evolutionary phase, following
a procedure quite similar to that used in investigating the evolutionary
properties of low mass He burning stars in galactic globular clusters.
For each assumption concerning the chemical 
composition and the age of a cluster,
 one obtains
from H-burning evolutionary computations  the mass of stars at the He ignition 
($M_{RG}$) and
both the He core mass ($M^{tip}_c$) and the amount of extrahelium 
$\Delta{Y}$
brought at the surface
by the first dredge up. When all these quantities are known,
Zero Age Horizontal Branch (ZAHB) models can be obtained, computing the
sequence of stellar structures, burning helium inside a He core of mass 
$M^{tip}_c$, surrounded by an envelope enriched by $\Delta{Y}$, with a
global mass
fulfilling the condition that $M_{tot}= M^{tip}_c+M_{envelope}\le{M}_{RG}$.
\par
\begin{figure}
\epsscale{.60}
\plotone{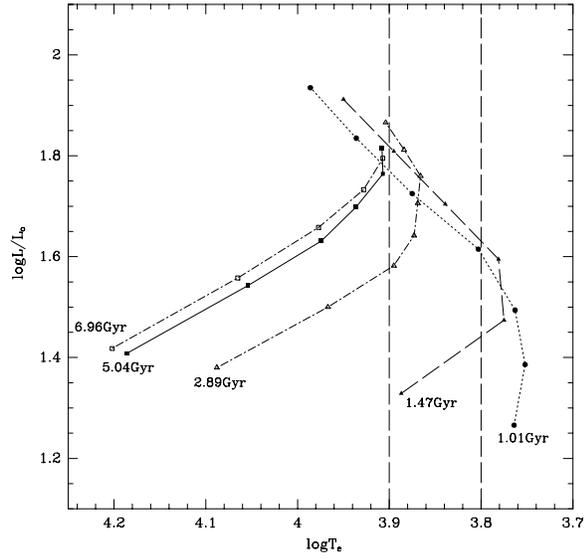}
\caption{The locus in the HR diagram of ZAHB structures for the various 
labeled assumptions about the cluster age. The mass of the various models 
is reported in Table 2.}
\end{figure}
However, at variance with the case of old globular cluster stars, 
the value of $M^{tip}_c$ is now sensitively depending on the value of $M_{RG}$, i.e. on the 
cluster age.
Since the luminosity of the ZAHB is largely dependent on the mass 
of the He core, one expects a large dependence of the ZAHB luminosity
on the age of the stellar system.
This occurrence is shown in figure 5, where ZAHB locations for the 
various labeled
assumptions about the cluster ages are plotted.
Table 2 gives selected evolutionary quantities for all the computed ZAHB models.
\par
%
%
%
%
%
%
\begin{deluxetable}{cccccc}
\scriptsize
\tablecaption{Selected evolutionary parameters for He burning stellar models.
The stellar mass, the luminosity and the effective temperature
of the ZAHB model and the central He burning lifetime (in $10^6yrs$)
are reported in the order for the various labeled assumptions 
about the cluster age, i.e., for the given mass
of the original progenitor.}
\tablehead{
\colhead{$M/M_{\odot}$} & \colhead{$log(L/L_{\odot})_{ZAHB}$} & 
\colhead{$log Te_{ZAHB}$} & \colhead{$\tau_{He-burn}$} & 
\colhead{$t_{HeF}(Gyr)$} & \colhead{$M_{pr}/M_{\odot}$}}
\startdata
0.60 &1.408 &4.186 &112 &5.04 &1.1 \nl
0.70 &1.543 &4.054 &103 &5.04 &1.1 \nl
0.80 &1.631 &3.974 &97  &5.04 &1.1 \nl
0.90 &1.699 &3.936 &93  &5.04 &1.1 \nl
1.00 &1.764 &3.907 &90  &5.04 &1.1 \nl
1.10 &1.816 &3.908 &88  &5.04 &1.1 \nl
\hline
0.60 &1.380 &4.088 &142 &2.89 &1.3 \nl
0.70 &1.500 &3.967 &130 &2.89 &1.3 \nl
0.80 &1.582 &3.895 &124 &2.89 &1.3 \nl
0.90 &1.642 &3.873 &119 &2.89 &1.3 \nl
1.00 &1.706 &3.869 &114 &2.89 &1.3 \nl
1.10 &1.760 &3.866 &110 &2.89 &1.3 \nl
1.20 &1.812 &3.884 &107 &2.89 &1.3 \nl
1.30 &1.866 &3.904 &104 &2.89 &1.3 \nl
\hline
0.60 &1.329 &3.887 &221 &1.47 &1.6 \nl
0.80 &1.474 &3.775 &193 &1.47 &1.6 \nl
1.00 &1.595 &3.781 &173 &1.47 &1.6 \nl
1.20 &1.704 &3.839 &158 &1.47 &1.6 \nl
1.40 &1.809 &3.895 &145 &1.47 &1.6 \nl
1.60 &1.912 &3.950 &132 &1.47 &1.6 \nl
\hline
0.60 &1.266 &3.764 &302 &1.01 &1.8 \nl
0.80 &1.386 &3.752 &263 &1.01 &1.8 \nl
1.00 &1.494 &3.763 &233 &1.01 &1.8 \nl
1.20 &1.615 &3.803 &209 &1.01 &1.8 \nl
1.40 &1.725 &3.875 &188 &1.01 &1.8 \nl
1.60 &1.835 &3.936 &170 &1.01 &1.8 \nl
1.80 &1.935 &3.986 &153 &1.01 &1.8 \nl
\enddata
\end{deluxetable}

As early recognized by Caloi,
Castellani \& Tornamb\'e (1978), 
inspection of figure 5 reveals that
increasing the total mass the effective temperature of a ZAHB model
decreases until a minimum temperature is reached, after that
the temperature starts increasing with mass. 
As for the origin of this minimum, one finds that 
for each given value of $M^{tip}_c$,
increasing the mass of the envelope, the temperature in the H burning
shell continuously increases whereas the density decreases.
The behavior of the central condition is almost specular, since
increasing the stellar mass the density increases and the temperature,
slightly, decreases. As a result,
the ratio between the luminosity due to H or He burning 
monotonously increases when 
the mass of the envelope is increased over the whole explored range of masses.
Thus the occurrence of the minimum in temperature cannot be
related to the relative efficiency of the burning.
The occurrence of this minimum can be much more simply
related to the evidence that increasing the stellar mass 
(i.e. the mass of the envelope),
the core and shell-burning regions 
behave more and more 
as a central energy source.
As a consequence, the more massive
ZAHB models shift towards
larger effective temperature approaching a MS-like location.
\par
Figs. 6 (a to d) show the evolutionary paths in the HR diagram
of the models in figure 5 during the phase of central
and shell He burning.
As already recognized for larger metallicity 
by Caputo \& Degl'Innocenti (1995),
one finds that the allowed range in magnitude of He burning red giants
increases when the cluster age decreases. Thus the observed range of
magnitudes 
can be used to put an upper limit to the 
age of stellar systems even in the $Z=10^{-5}$ case.
As for the post-HB evolution, all computed models have envelopes massive
enough to make them approach their Hayashi tracks during the double shell
burning phase.
\par
\begin{figure}
\epsscale{.60}
\plotone{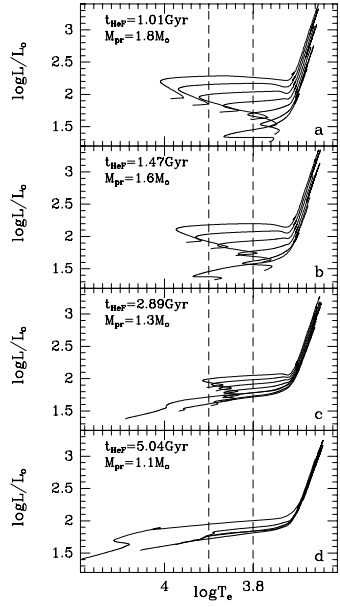}
\caption{The evolutionary path in the HR diagram of He burning models
for the labeled assumptions about the cluster age. The initial mass
of the progenitor ($M_{pr}$) is also reported. The vertical dashed lines
sketch the temptative location of the strip for pulsational instability.}
\end{figure}
In both Figures 5 and 6 (a to d) we report  
the temptative location of the region for pulsational
instability,
allowing a discussion of the possible occurrence of 
variable stars in stellar populations with Z=$10^{-5}$.
As well known, old very metal poor systems cannot produce
ZAHB pulsators since the ZAHB locations are in all cases  
hotter than the
instability strip. According to Cassisi et al. (1995), this is the case
for metallicities ${\log}Z\le-5$ and ages larger than about 7 Gyrs.
As expected, now one finds
that, for ages smaller than - about - 
5.1 Gyrs metal deficient clusters start allowing 
ZAHB pulsators.
Decreasing the cluster age, the range of masses in the instability
strip increases.
For ages of about 2.8-2.9 Gyrs
one finds that all
ZAHB models more massive than $0.8M_\odot$ 
are inside the instability strip.
In similar clusters one should expect an
anomalous clump of He burning pulsating stars. 
However, a detailed discussion on the pulsational scenario 
concerning similar
metal poor variables is beyond the aim of the present work. For a deeper investigation
on the pulsational properties of these metal poor stars, we will address to a 
forthcoming paper (Bono et al. 1996).
\par
Figures 7 (a to d) and 8 (a to d) show the time evolution of luminosity
and effective temperature
of the models in figures 6 (a to d) 
during both central and shell He burning phases. 
\begin{figure}
\epsscale{.60}
\plotone{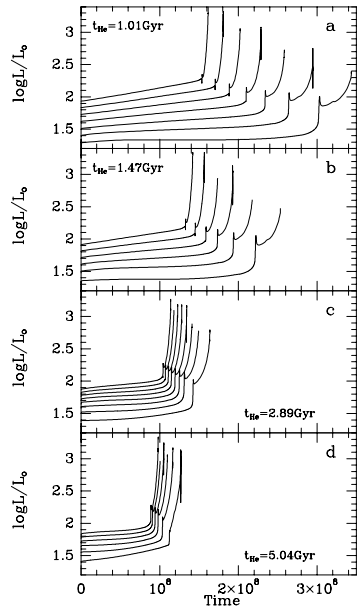}
\caption{The behavior with the time of the luminosity during the central
and shell He burning phases for the labeled assumptions concerning the cluster
ages.}
\end{figure}
\begin{figure}
\epsscale{.60}
\plotone{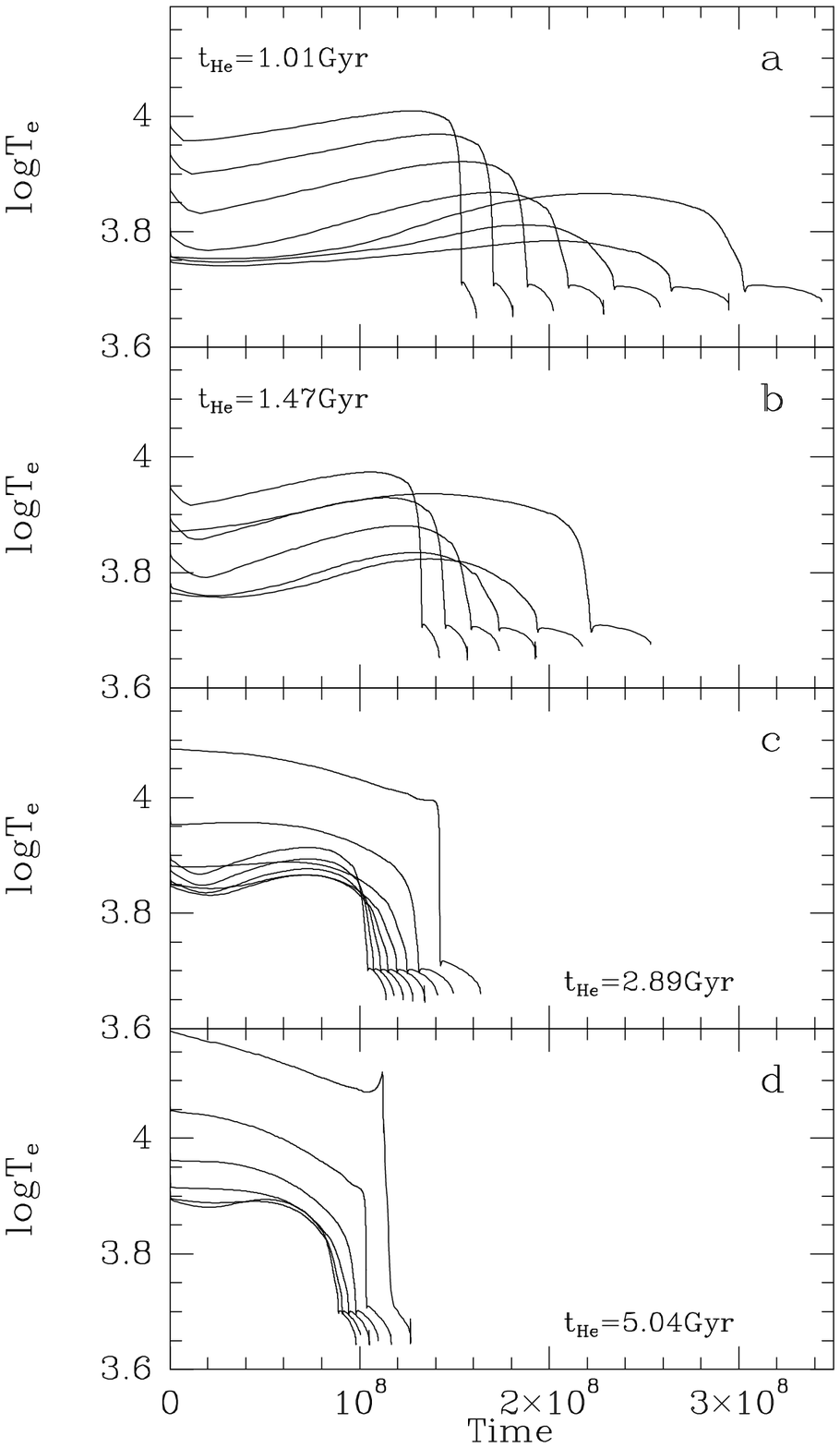}
\caption{The behavior with the time of the effective temperature for all models
computed in the present work during the central
and shell He burning phases for the labeled assumptions concerning the cluster
ages.}
\end{figure}
Without enter into a detailed discussion, let us only notice the
progressive increase of the He central burning lifetimes when the cluster age
is decreased.
This is the expected result of the corresponding decrease
of the mass of the He-core in ZAHB models. 
As a consequence, one expects
an increasing evidence for He-burning giants which will eventually
dominate the cluster giant population.

\section{Conclusions.}
\par
This paper investigates the evolutionary properties
of relatively massive, metal deficient stellar structures.
The first part of the investigation has been devoted to 
H burning stars, presenting selected isochrones 
for ages in the range 800 Myr - 7 Gyr, increasing the range
of ages covered by previous investigations. 
 
Selected sets of HB models have been computed
under different assumptions about the ages of the stellar system.
We confirm the results already given for larger stellar metallicities
about the possible, and sometime probable, 
occurrence of anomalous, overluminous variable stars,
in relatively young, metal deficient system.
\par
Both evolutionary tracks and isochrones are available by
electronic mail upon request to cassisi@astrte.te.astro.it.
\acknowledgments

We thank F. Caputo for helpful discussions and for suggesting
the need for such a completion of the current evolutionary scenario. S.C.
thanks also O. Straniero and A. Tornamb\'e for useful and stimulating 
discussions.

\end{document}